\documentclass[conference]{IEEEtran}
\IEEEoverridecommandlockouts
\usepackage{cite}
\usepackage{amsmath,amssymb,amsfonts}
\usepackage{algorithmic}
\usepackage{graphicx}
\usepackage{textcomp}
\usepackage{xcolor}
\usepackage[caption=false,font=normalsize,labelfont=rm,textfont=rm]{subfig}
\usepackage{array}
\usepackage{stfloats}
\usepackage{url}
\usepackage{verbatim}
\usepackage{multirow}
\usepackage{caption}
\usepackage{hyperref} 
\usepackage{makecell}
\usepackage{bm}

\def\BibTeX{{\rm B\kern-.05em{\sc i\kern-.025em b}\kern-.08em
    T\kern-.1667em\lower.7ex\hbox{E}\kern-.125emX}}

\begin{document}

\title{Adaptive Wireless Image Semantic Transmission and Over-The-Air Testing
}

\author{\IEEEauthorblockN{Jiarun Ding\IEEEauthorrefmark{1}, 
Peiwen Jiang\IEEEauthorrefmark{1}, Chao-Kai Wen\IEEEauthorrefmark{2},
and Shi Jin\IEEEauthorrefmark{1}}
\IEEEauthorblockA{\IEEEauthorrefmark{1}National Mobile Communications Research Laboratory, Southeast University\\ 
Nanjing 210096, China, Email: \{jrunding, peiwenjiang, jinshi\}@seu.edu.cn}
\IEEEauthorblockA{\IEEEauthorrefmark{2}Institute of Communications Engineering, National Sun Yat-sen University\\
Kaohsiung 80424, Taiwan, Email: chaokai.wen@mail.nsysu.edu.tw.}}

\maketitle

\begin{abstract}
Semantic communication has undergone considerable evolution due to the recent rapid development of artificial intelligence (AI), significantly enhancing both communication robustness and efficiency. Despite these advancements, most current semantic communication methods for image transmission pay little attention to the differing importance of objects and backgrounds in images. To address this issue, we propose a novel scheme named ASCViT-JSCC, which utilizes vision transformers (ViTs) integrated with an orthogonal frequency division multiplexing (OFDM) system. This scheme adaptively allocates bandwidth for objects and backgrounds in images according to the importance order of different parts determined by object detection of you only look once version 5 (YOLOv5) and feature points detection of scale invariant feature transform (SIFT). Furthermore, the proposed scheme adheres to digital modulation standards by incorporating quantization modules. We validate this approach through an over-the-air (OTA) testbed named intelligent communication prototype validation platform (ICP) based on a software-defined radio (SDR) and NVIDIA embedded kits. Our findings from both simulations and practical measurements show that ASCViT-JSCC significantly preserves objects in images and enhances reconstruction quality compared to existing methods.
\end{abstract}


\section{Introduction}
The future sixth-generation (6G) framework propels the application of AI in communications \cite{1}. Benefit from the remarkable semantic extraction capabilities of AI, semantic communication is gradually brought to the forefront of research \cite{61}. Semantic communication is dedicated to ensuring reliable semantics transmission \cite{5} and typically leverages autoencoder-based joint source-channel coding (JSCC) architecture for multimedia data transmission. This differs from separate source and channel coding based on Shonnon’s information theory, which is optimal for a memoryless source and channel when the code length are not constrained. However, infinite code length is impractical, so separation-based scheme is typically suboptimal, making JSCC-based semantic communication competitive in several scenarios.

Deep learning based JSCC for semantic communication is prevalent currently and it has been extensively studies in various domains such as text, speech, image and video \cite{16,13,24}. Images and videos typically contain more redundant information, making JSCC-based semantic communications more challenging. DeepJSCC \cite{14} uses deep neural networks (NNs) for image transmission. In \cite{34}, the proposed scheme adaptively controls transmission rate according to channels and source features. OFDM and multipath fading channels are considered in \cite{36}. In \cite{37}, an ingenious NN utilizing channel state information (CSI) is proposed to implement channel-adaptive OFDM system. Furthermore, multiple-input multiple-output (MIMO) and attention mechanisms are also employed to enhance throughput and implement channel adaptation \cite{38,39}. To conform to digital modulation standards, \cite{43} proposes DeepJSCC-Q, leveraging a fixed channel input constellation and achieving similar performance to unquantified DeepJSCC \cite{14}. However, better performance can only be achieved when the modulation order is high. The aforementioned works have given little consideration to changing requirements, such as prioritizing objects in images over backgrounds under resource-constrained conditions. Our study focuses on offering a method to solve this issue, making the JSCC-based scheme more competitive.

As a bridge connecting simulations and practical applications, OTA testing is required for the proof-of-concept of intelligent communication including semantic communication. The authors in \cite{46} deploy AI-aided online adaptive OFDM receiver using C/C++ on RaPro \cite{45}. Since intelligent algorithms rely on C/C++, it is difficult for quick deployment. A testbed based on host personal computer (PC) and SDR specially targeting wireless semantic communication is proposed in \cite{48}. While the symbols are full-resolution and it is still a single-carrier transmission, there is a significant difference between this approach and modern wideband-based wireless communications. Therefore, testbeds need to be further improved and studied.

Inspired by aforementioned issues regarding to algorithms and testbeds, we conducted this study, and the major contributions of this paper can be summarized as follows:
\begin{enumerate}
    \item We propose a novel adaptive wireless image semantic transmission scheme named ASCViT-JSCC, which complies with digital modulation standards. ASCViT-JSCC focuses on preserving objects in images and achieves a trade-off between objects and backgrounds based on channel environments.
    \item We designed a testbed named ICP based on a USRP-2943R and NVIDIA embedded kits to establish a hardware foundation for research on intelligent communication. By deploying the proposed scheme and comparative schemes, we verify the superiority of the testbed and algorithms and present practical measurements.
\end{enumerate}
\begin{figure*}[!t]
	\vspace{-4mm}
    \centerline{\includegraphics[width=5in]{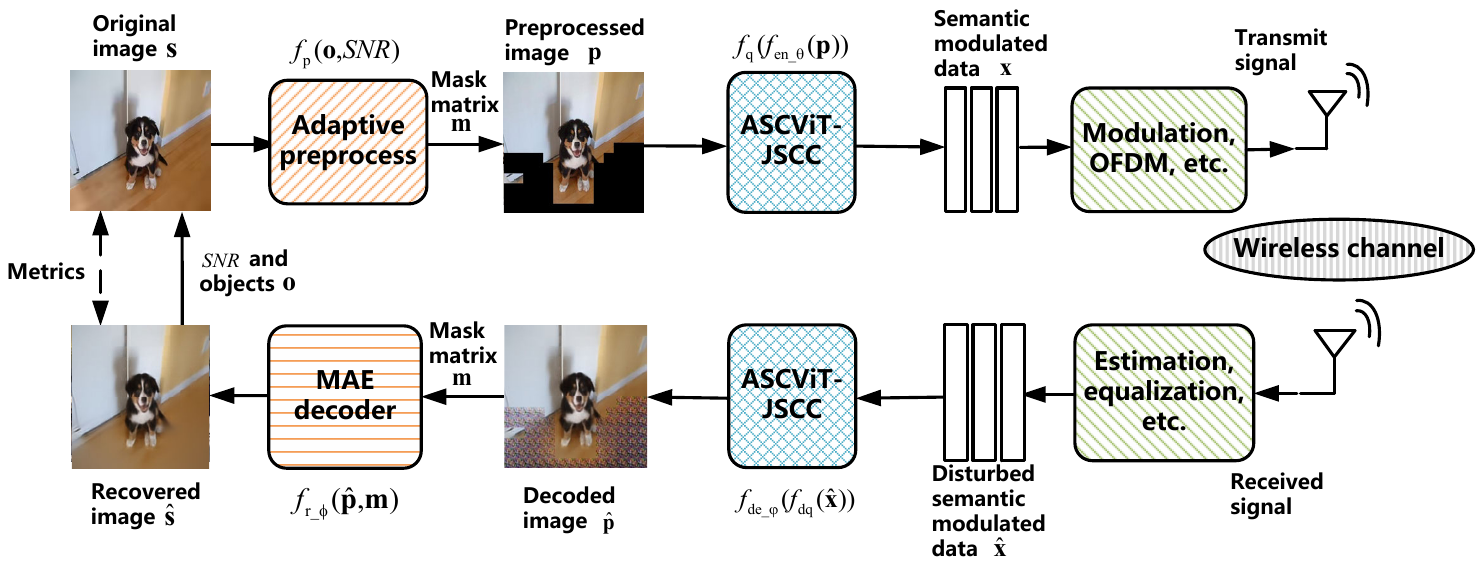}}
	\caption{The structure of ASCViT-JSCC.}
	\label{fig1}
    \vspace{-4mm}
\end{figure*}

\section{SYSTEM MODEL AND TRANSCEIVER DESIGN}
\subsection{System Model}
The structure of ASCViT-JSCC is depicted in Fig. \ref{fig1}. Notate the CSI and objects from receiver as $\textit{SNR}$ and $\textbf{o}\in{\mathbb{R}^{1\times{O}}}$, which refer to signal-to-noise (SNR) and objects, respectively. $O$ is the number of objects. A mask matrix is generated according to $\textbf{o}$ and $\textit{SNR}$ by
\begin{equation}
\textbf{m}=f_{\rm{p}}(\textbf{o}, \textit{SNR}), \textbf{m}\in{\mathbb{R}^{H\times{W}\times{C}}}, \label{eq0} \end{equation}
where $H$, $W$, $C$ and $f_{\rm{p}}(\cdot)$ denote height, width, pixel channel number of image and mask matrix generating function, respectively.

Notate the original image as $\textbf{s}\in{\mathbb{R}^{H\times{W}\times{C}}}$. $\textbf{s}$ is processed by $\textbf{s}\odot\textbf{m}$ to generate the preprocessed image $\textbf{p}\in{\mathbb{R}^{H\times{W}\times{C}}}$, where $\odot$ represents the element-wise product. Then $\textbf{p}$ is encoded by JSCC NN $f_{\rm{en\_{\rm{\bm{\theta}}}}}(\cdot)$ to generate semantic floating-point data. Quantization module $f_{\rm{q}}(\cdot)$ is introduced to further quantify floating-point data to modulated data $\textbf{x}\in{\mathbb{R}^{1\times{N}}}$, where $\textit{N}$ is the number of modulated data. The above process can be expressed as
\begin{equation}
\textbf{x}=f_{\rm{_q}}(f_{\rm{en}\_{\rm{\bm{\theta}}}}(\textbf{p})). \label{eq1} 
\end{equation}

Through physical-level signal processing, semantic modulated data becomes OFDM symbols and are transmitted to the receiver over wireless channel. All the received symbols restore to disturbed semantic modulated data $\hat{\textbf{x}}\in{\mathbb{R}^{1\times{N}}}$ through channel estimation and equalization, etc. Afterwards, dequantization module $f_{\rm{dq}}(\cdot)$ and JSCC decoding NN $f_{\rm{de}\_{\rm{\bm{\varphi}}}}(\cdot)$ decode distributed semantic modulated data to image $\hat{\textbf{p}}\in{\mathbb{R}^{H\times{W}\times{C}}}$, whose backgrounds are still masked. Finally, the pretrained model masked autoencoder (MAE) \cite{51} $f_{\rm{r}\_{\rm{\bm{\phi}}}}(\cdot)$ is utilized to recover the backgrounds and produce the final recovered image $\hat{\textbf{s}}$. The above process can be expressed as
\begin{equation} \hat{\textbf{s}}=f_{\rm{r}\_{\rm{\bm{\phi}}}}(f_{\rm{de}\_{\rm{\bm{\varphi}}}}(f_{\rm{dq}}(\hat{\textbf{x}})),\textbf{m}). \label{eq3} 
\end{equation}

\begin{figure}[htbp]
	\centerline{\includegraphics[width=3in]{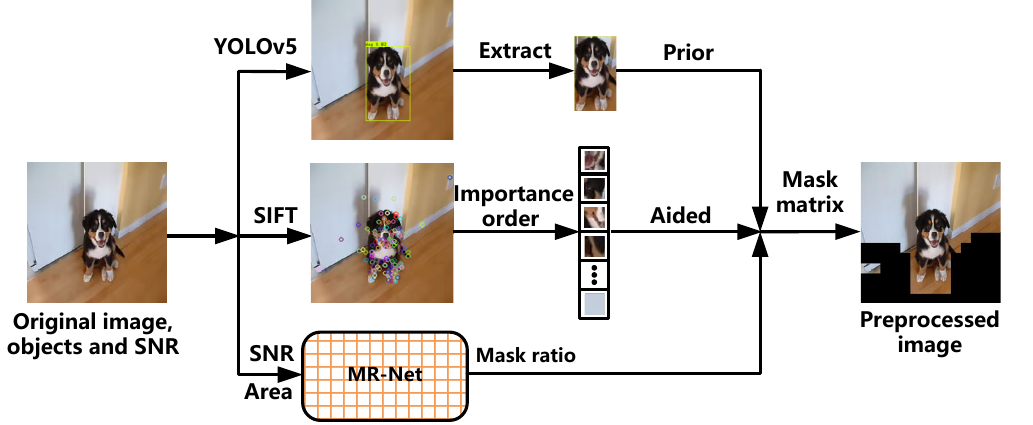}}
	\caption{The adaptive preprocessing.}
	\label{fig2}
    \vspace{-7mm}
\end{figure}
\subsection{Adaptive Preprocess}
Adaptive preprocessing module adapts to different SNRs and user requirements which is showed in Fig. \ref{fig2}.
\begin{figure*}[!t]
	\centerline{\includegraphics[width=5.0in]{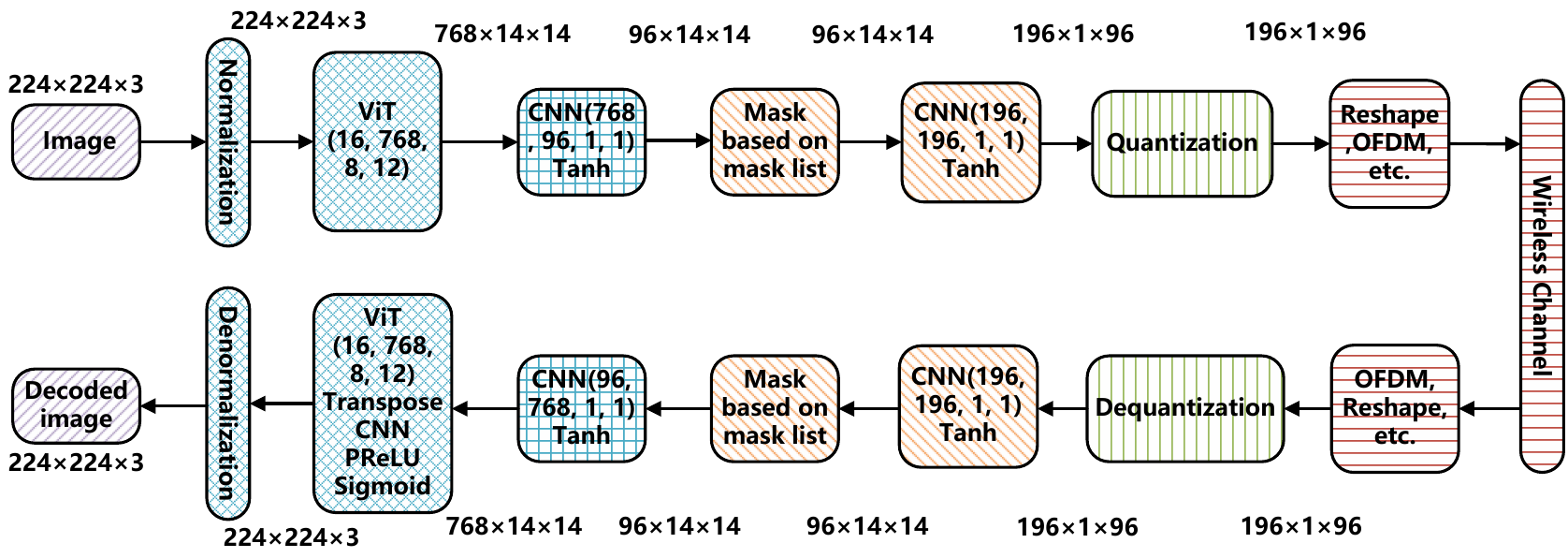}}
	\caption{The structure of ASCViT-JSCC NN. For ViT(a, b, c, d), a, b, c and d denote patch size, embedded dimension, head number and block number, respectively. For CNN (a, b, c, d), a, b, c and d respectively denote input channel number, output channel number, kernel size and stride. The numbers labelled next to the graph are output dimension of networks.}
	\label{fig3}
    \vspace{-4mm}
\end{figure*}
Masked patches serve as redundant information to safeguard unmasked patches during transmission. Therefore, under varying wireless channel conditions, ASCViT-JSCC must dynamically adjust mask ratio (MR) to achieve a trade-off between information and redundancy.

Input image size is $224\times{224}\times{3}$, representing height, width and pixel channel number, respectively. We split an image into 196 patches, each with a size of $16\times{16}$. YOLOv5 is leveraged to generate detection results. Patches that belong to objects are assigned the highest priority. Additionally, SIFT detects the feature points and calculates the number of feature points of each patch. Based on feature points number, the patches are sorted. Finally, this process results in an importance order of the 196 patches.

In Fig. \ref{fig2}, MR-Net is depicted as a fully-connected NN responsible for determining the MR based on the SNR and the area of objects. When SNR is higher and area of objects is smaller, MR-Net outputs a lower MR. This ensures that more information patches are retained for encoding and transmission over wireless channel. Conversely, a higher MR indicates that more patches are masked as redundant information to preserve unmasked patches. Combing the MR with the importance order, ASCViT-JSCC generates the optimal mask list, represented as a binary sequence with the length of 196. Then a mask matrix consisting of binary values is produced according to the mask list and input image is masked by it to generate the preprocessed image.
\subsection{ASCViT-JSCC NN Structure}
The structure of ASCViT-JSCC NN is illustrated in Fig. \ref{fig3}. Firstly, the NN normalizes the pixel value of input image to $[0, 1]$. Then, ViT extracts semantic information and compression convolutional neural network (CNN) compresses it as much as possible. We design a mask operation followed by a channel coding CNN to leverage masked patches to help recover unmasked patches. Next, quantization module quantifies the floating-point data into semantic modulated data selected from $[-3, -1, 1, 3]$, which corresponds to 16-ary quadrature amplitude modulation (16QAM).
Lastly, signal processes in physical layer and wireless channel are brought into the ASCViT-JSCC NN. Networks at the receiver are reverse of those at the transmitter. All designs are differentiable to enable end-to-end optimization.

We solely optimize the parameters of ViTs and compression CNNs with OFDM in noiseless channel. The loss function of this stage is mean square error (MSE) which can be expressed as
\begin{equation}MSE(\textbf{p}, \hat{\textbf{p}})=\frac{1}{H\times{W\times{C}}}\sum_{i=1}^{H\times{W\times{C}}}(p_i-\hat{p_i})^2, \label{eq6} \end{equation}
where $p_i$ and $\hat{p_i}$ denotes the $i$-th pixel channel value. In experiments, we determine compression limit by ensuring average peak signal-to-noise ratio (PSNR) of decoded image higher than 38dB. This results in a confirmatory bandwidth compression ratio (BCR) \cite{14} equal to $\frac{1}{16}$. Thus the number of symbols transmitted over wireless channel is 9408. By minimizing the MSE between the preprocessed image $\textbf{p}$ and the decoded image $\hat{\textbf{p}}$, ViTs and compression CNNs are optimized for efficiently accomplishing pixel-level image reconstruction task.

Then, the parameters of ViTs and compression CNNs are frozen, and channel coding CNNs, etc. are adopted to networks. The entire NNs are trained in additive white Gaussian noise (AWGN) and Rayleigh fading channels, but only the parameters of channel coding CNNs are optimized during this stage. The loss function for this stage is
\begin{equation}
MSE_{\rm{um}}(\textbf{p}, \hat{\textbf{p}})=\frac{1}{H\times{W\times{C}}}\sum_{i=1}^{H\times{W\times{C}}}(p_i-(\hat{p}_i \cdot m_i))^2\times{\frac{N_{\rm{T}}}{N_{\rm{U}}}}, \label{eq7} \end{equation}
where $N_{\rm{T}}$, $N_{\rm{U}}$ and $m_i$ denote the total patch number, unmasked patch number and the $i$-th value of the mask matrix, respectively. Finally, we fine-tune all parameters of networks in different channel environments.

%%%%%%%%%%%%%%%%  3 SIMULATION RESULTS %%%%%%%%%%%%%%%%%%%%%%
\section{Numerical RESULT}
This section shows the numerical results and analysis of the proposed ASCViT-JSCC. Specifically, all NNs are trained on NVIDIA Tesla V100 graphics processing unit (GPU) 32G and we test the performance on NVIDIA GTX 1650 GPU 4G because its computational power is similar to that of the proposed testbed.
\subsection{Experimental Setup}
All experiments are conducted on ImageNet2012 \cite{56}. We randomly select 20,000 images as training set and 1,000 images as validation and test set, respectively. The OFDM system comprises 64 subcarriers, with 55 data subcarriers and others for pilot. Least squares (LS) channel estimation and zero forcing (ZF) signal detection are adopted and the utilized modulation is 16QAM. AWGN and Rayleigh fading channels are considered. The training batch size is set to 8 and the number of epoch is 200. Adam optimizer with learning rate 0.0002 is adopted in training. All networks are trained at 10dB SNR, as well as SNRs uniformly sampled from the range of $[-5, 15]$.

Two baselines are considered to compare with the proposed scheme. (1) Better portable graphics (BPG) \cite{58} and low-density parity-check codes (LDPC) \cite{62}. 1440 code length and one-half code rate LDPC coding is leveraged. (2) CNN-based DeepJSCC \cite{14} with the same quantization modules used in ASCViT-JSCC. The number of modulated symbols of two baselines is controlled approximately equal to 9408.

To facilitate the evaluation of different schemes, we combine PSNR, structure similarity index measure (SSIM) and confidence score (CS) into two metrics: PSNR+CS and SSIM+CS. PSNR and SSIM are prevalent metrics to evaluate the similarity of two images. CS originates from YOLOv5 and represents the quality of objects. Specifically, we normalize the PSNR to $(0,1)$ by dividing it by 40 and then combine it with CS by
\begin{equation}PSNR+CS=\frac{PSNR}{40}\times{\frac{1}{2}}+CS\times\frac{1}{2}. \label{eq8} \end{equation}
Likewise, SSIM+CS can be calculated by
\begin{equation}SSIM+CS=SSIM\times{\frac{1}{2}}+CS\times\frac{1}{2}, \label{eq9} \end{equation}
where $\frac{1}{2}$ is the weight of two metrics. The reason for this set is the quality of objects and reconstructed images are equally important in this study.
\begin{figure} [htbp]
    \vspace{-5mm}
	\centering  %居中
    \subfloat[PSNR+CS versus MR]{
        \centering  %居中
        \includegraphics[scale=0.48]{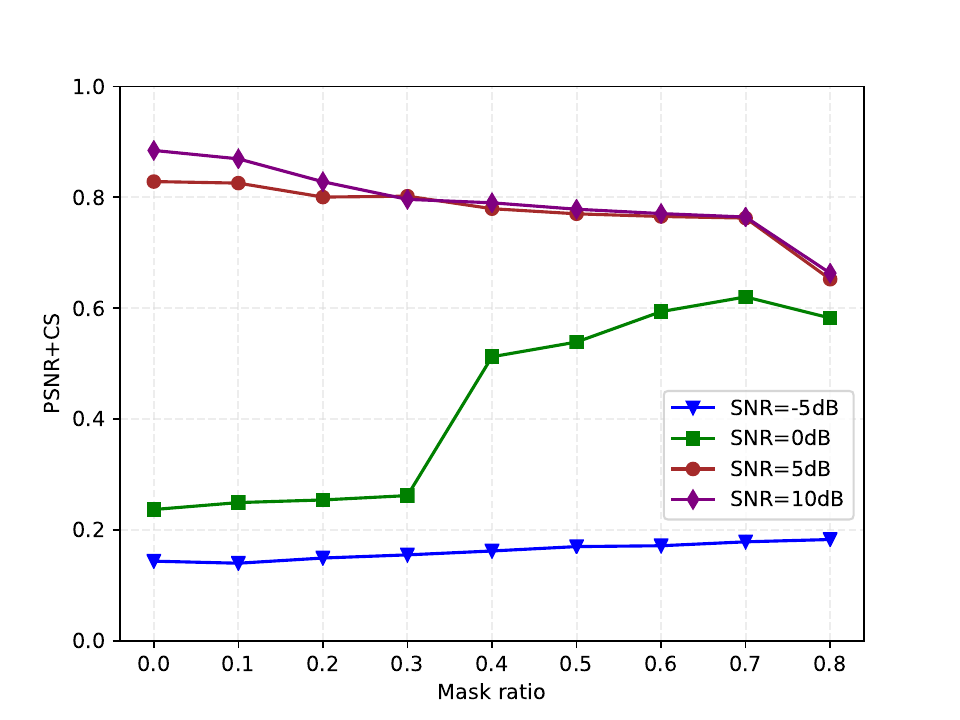} 
        \label{fig4:a}
    }
    \vspace{-4mm}
    \  % 两个子图上下排序
	\subfloat[SSIM+CS versus MR]{
        \centering  %居中
        \includegraphics[scale=0.48]{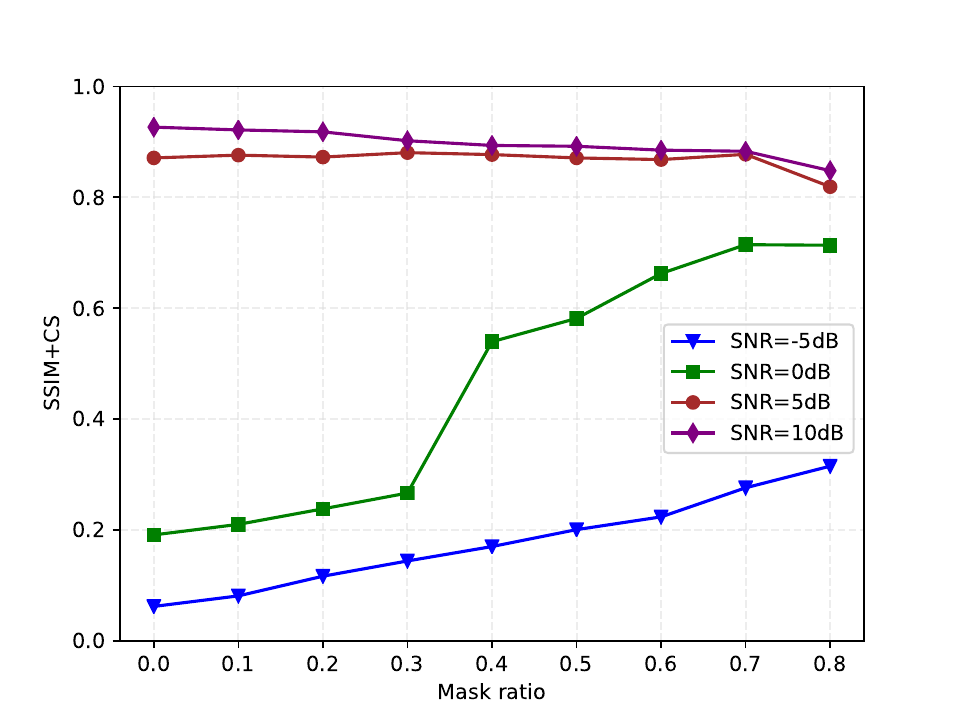} 
        \label{fig4:b}
    }
	\caption{Two metrics versus MR. SNR indicates the SNRs of test channels.}
    \label{fig4}
    \vspace{-4mm}
\end{figure}

\subsection{Performance versus MR}
The network trained at 10dB in AWGN channel is selected to evaluate performance versus MR to confirm the optimal MR at different SNRs. The results are shown in Fig. \ref{fig4}.
\begin{figure*} [!h]
	\centering
	\subfloat[PSNR+CS versus SNR AWGN]{
		\includegraphics[width=3in]{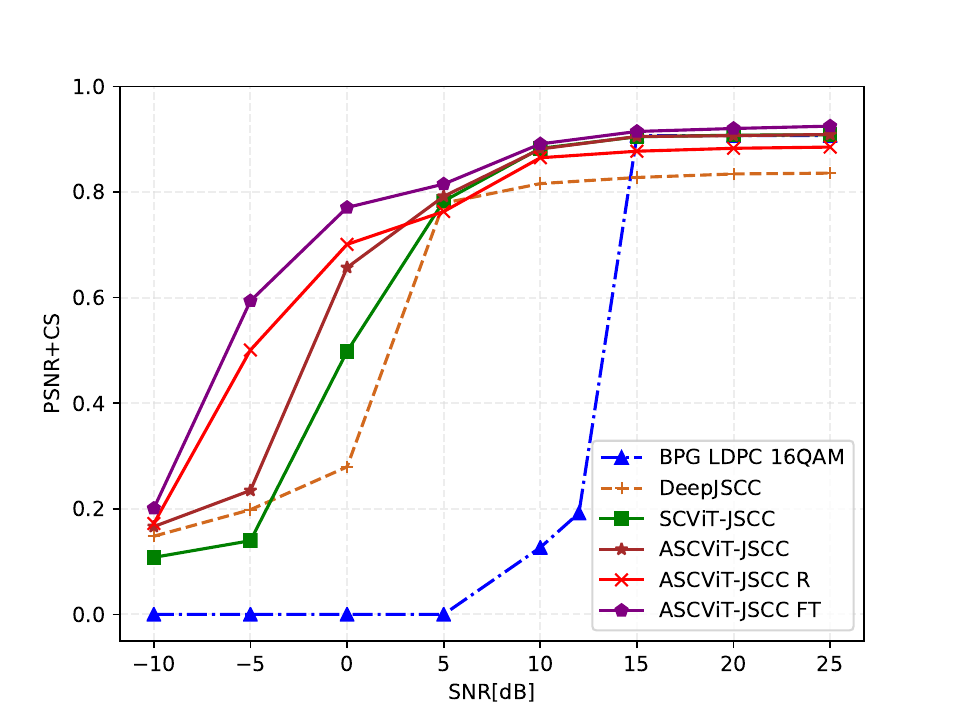}
        \label{fig5:a}
    }
	\subfloat[SSIM+CS versus SNR AWGN]{
		\includegraphics[width=3in]{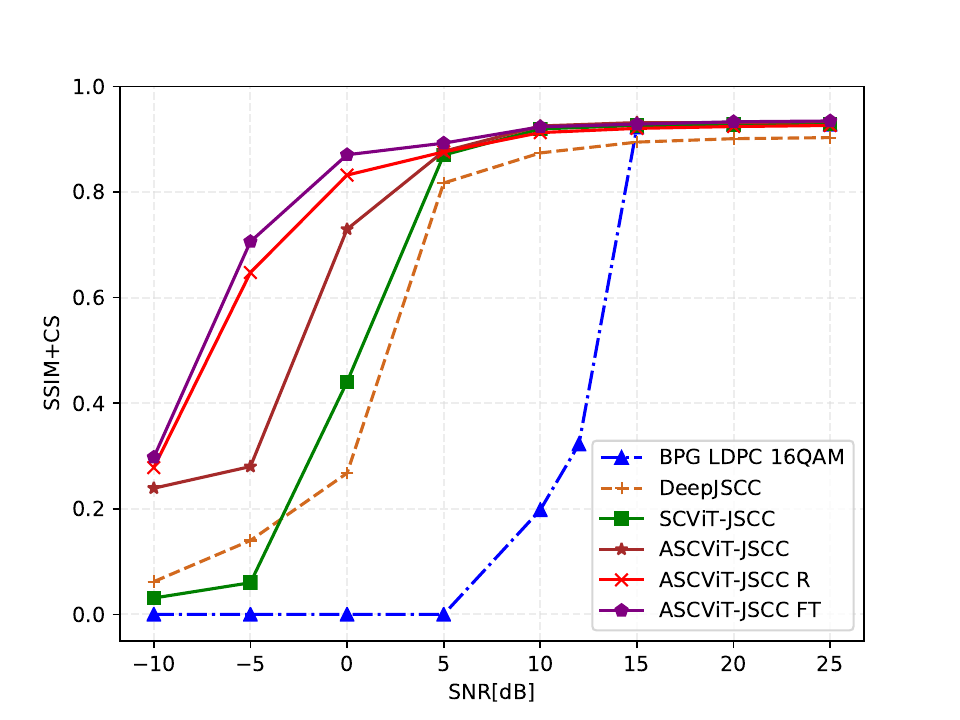}
        \label{fig5:b}
    }
    \\
    \vspace{-4mm}
    \subfloat[PSNR+CS versus SNR Rayleigh]{
		\includegraphics[width=3in]{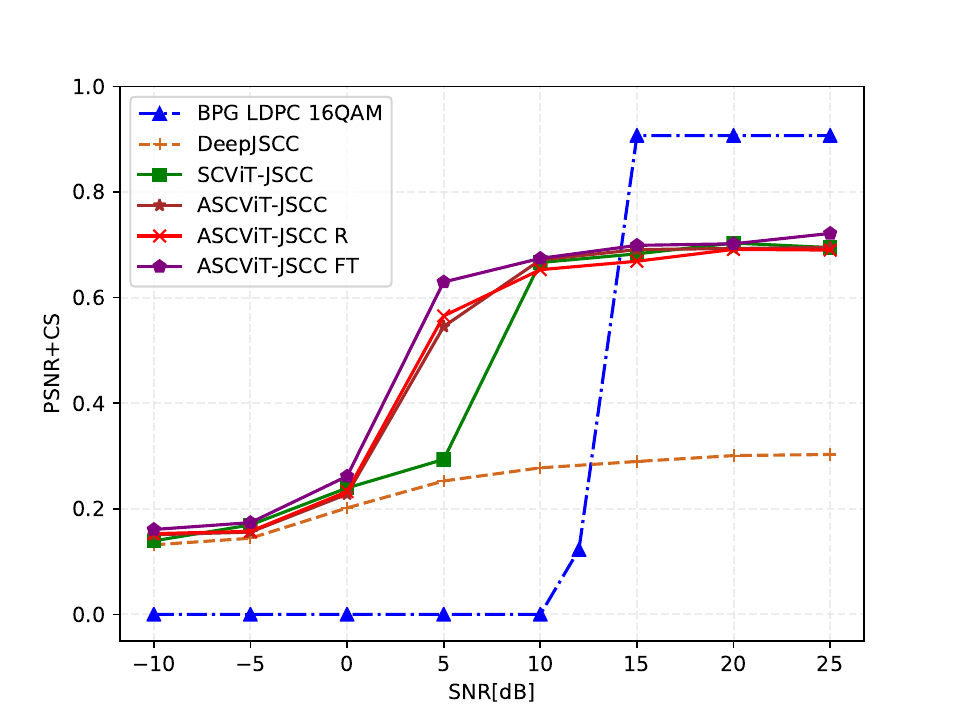}
        \label{fig5:c}
    }
    \subfloat[SSIM+CS versus SNR Rayleigh]{
		\includegraphics[width=3in]{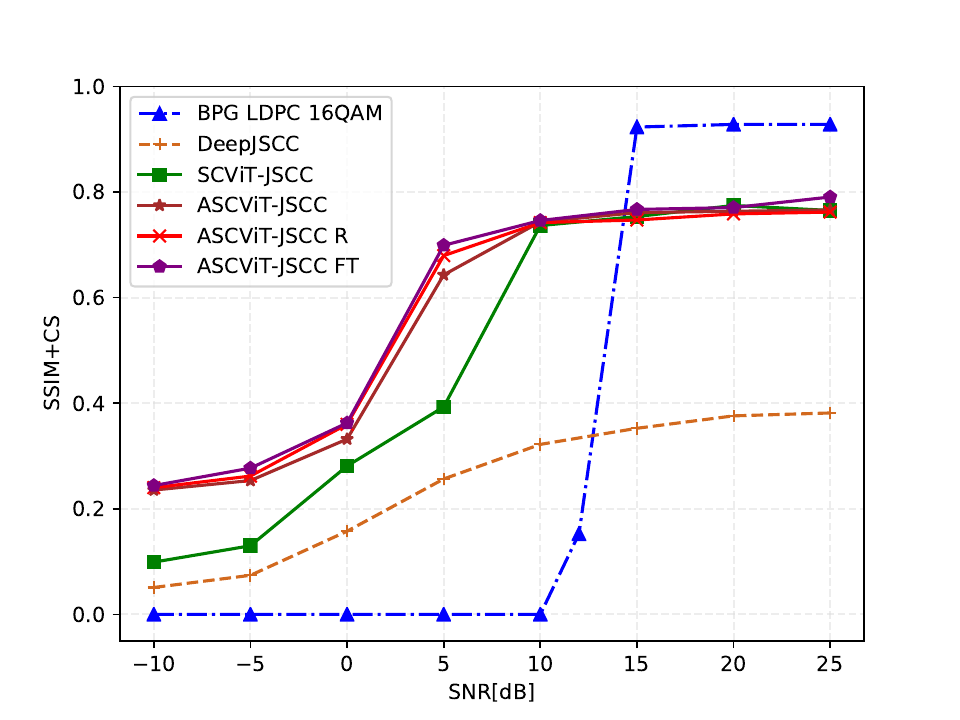}
        \label{fig5:d}
    }
	\caption{Performance of ASCViT-JSCC compared to other schemes in AWGN and Rayleigh fading channels. ``R'' indicates that NNs are trained at ramdom SNRs uniformly sampled from [-5, 15]. ``FT''  indicates the networks are fine-tuned at test SNRs.}
	\label{fig5}
    \vspace{-6mm}
\end{figure*}

As illustrated in Fig. \ref{fig4}, both PSNR+CS and SSIM+CS at 0dB exhibit wide variations across different MRs. However, metrics under other conditions vary slowly. It is observed that metrics decreases as MR increases at both 5dB and 10dB SNRs. This trend indicates that under these conditions, the quality of reconstructed images is excellent enough, rendering additional mask operations unnecessary. Conversely, at 0dB SNR, metrics increases as MR increases, especially from 0.3 to 0.4. While when the MR increases from 0.7 to 0.8, metric decrease instead, suggesting that a MR of 0.7 is optimal under this condition. Excessive masking may lead to the masking of patches that could otherwise be better recovered. At low SNRs such as -5dB, performance consistently improves as MR increases. This phenomenon validates the concept of leveraging masked patches to help recover unmasked patches. Finally, based on observations in Fig. \ref{fig4}, it can be concluded that PSNR+CS is more sensitive to MR in relatively high SNR regime, whereas in low SNR regime, it behaves conversely. SSIM+CS demonstrates the opposite behavior to PSNR+CS with respect to MR. 

The optimal MRs can be obtained at different SNRs. For instance, under -5dB SNR condition, a MR of 0.8 is most suitable for improving system performance, while a ratio of 0.7 is optimal at 0dB SNR. Considering the size of objects in images, we set the MR in the range of $[0, 0.7]$ to prevent MR-Net from masking objects which could result in worse performance instead. Consequently, a training dataset is constructed based on the aforementioned results and MR-Net is trained on it. With a refined MR-Net network structure, ASCViT-JSCC can fully showcase its performance. In subsequent experiments, the structure and parameters of MR-Net are fixed to ensure the overall system performance.

\subsection{Performance of ASCViT-JSCC in AWGN and Rayleigh Fading Channels}

Fig. \ref{fig5} present the numerical results at different SNRs in AWGN and Rayleigh fading channels. In AWGN channels, the traditional scheme increases sharply when SNR ranges from 10dB to 15dB. However, in lower SNR regime, BPG is unable to decode the high bit error rate bits, resulting in no performance improvements in higher SNR regime. This is the limitation of the traditional scheme, known as the \textit{cliff effect}. In contrast, schemes based on NNs exhibit smoother curves and effectively mitigate this issue. SCViT-JSCC outperforms DeepJSCC at most SNRs due to its global perception. In low SNR regime, ASCViT-JSCC which is SCViT-JSCC with channel adaptive masking operation outperforms SCViT-JSCC. However, they have the same performance in high SNR regime. As channel condition improves, MR-Net outputs a 0 MR, resulting in ASCViT-JSCC and SCViT-JSCC exhibiting the same performance. The parameters obtained by training at random SNRs outperform those obtained from fixed SNRs in low SNR regime. This is because during the training phase, networks have experienced harsh channel conditions, leading to better performance in low SNR regime during the test phase.

In Rayleigh fading channel, the curve of traditional scheme shifts to the right compared to that in AWGN channel. However, the optimal performance of traditional scheme remains unchanged. ASCViT-JSCC still outperforms SCViT-JSCC and DeepJSCC. However, due to the difficulty of training in Rayleigh fading channel, ASCViT-JSCC trained at fixed SNR and random SNRs approximately performs at the same level. Only fine-tuning at a specific SNR can result in a slight performance improvement. The numerical results of DeepJSCC are extremely low compared to those in AWGN channels because YOLOv5 cannot detect objects in vast majority of images. In addition, it is obvious in Rayleigh fading channel than AWGN channel that the performance of ASCViT-JSCC increases relatively sharply from 0dB to 5dB. Although NN based schemes do not perform well in high SNR regime, they still have advantages in low SNR regime due to their capability to mitigate \textit{cliff effect}.

In general, our proposed scheme ASCViT-JSCC outperforms DeepJSCC with quantization modules in all SNR regimes and performs on par with the traditional scheme, even in very high SNR regimes. Similar to other schemes based on NNs, ASCViT-JSCC can mitigate the \textit{cliff effect}, thereby improving the quality of received images under extremely poor channel conditions. 
\section{OTA TEST}
\subsection{System Components and Parameters}
\begin{figure}[!h]
    \centering
	\includegraphics[width=3.2in]{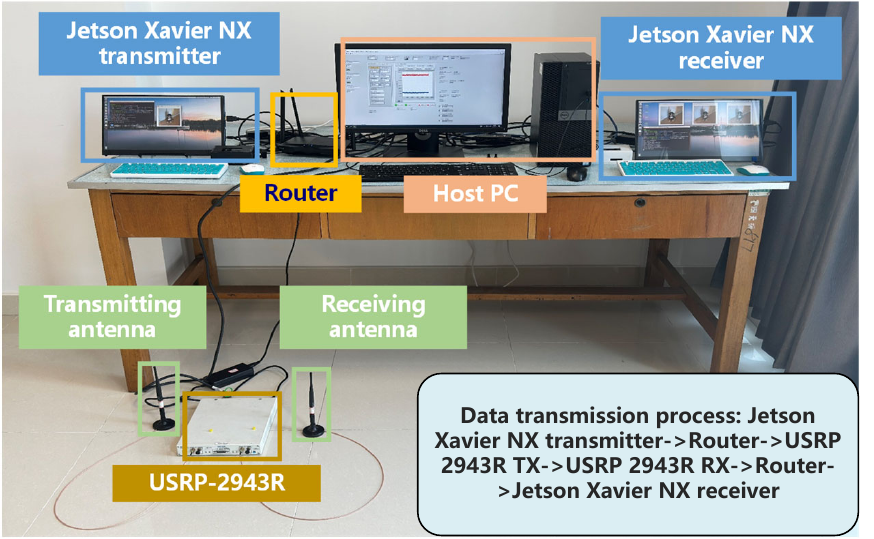}
    \caption{System components of ICP.}
	\label{fig6}
    \vspace{-7mm}
\end{figure}
% \setlength{\belowcaptionskip}{-5cm} 

%%%%%%%%%%%%%%%%  4 PROTOTYPE VALIDATION %%%%%%%%%%%%%%%%%%%%%%
Fig. \ref{fig6} shows the system components of ICP. ICP is mainly composed of two Jetson Xavier NXs, a USRP-2943R with two antennas, a host PC and a router. Linux runs on Jetson Xavier NX, and so we can build any deep learning environments supporting ARM64 architecture. USRP-2943R consists of two radio frequency (RF) transceivers of 120 MHz bandwidth. In our study, we leverage one RF channel to transmit modulated radio signals and another to receiver, to implement self-transmitting and self-receiving in one USRP-2943R. The router forms Ethernet for user datagram protocol (UDP) transmitting and receiving. The data transmission process are shown in the bottom of Fig. \ref{fig6} and the parameters of ICP is presented in TABLE \ref{tab1}.

\subsection{Experimental Results and Performance Analysis}
Simple scenarios are considered by varying distance between two antennas. The practical measurements are presented in TABLE \ref{tab2}. It is observed that ASCViT-JSCC, trained at random SNRs, outperforms other schemes in medium and low SNR regimes. In high SNR regime, the traditional scheme still exhibits the optimal performance. Nevertheless, ASCViT-JSCC trained at a fixed 10dB SNR falls just slightly short of the traditional scheme in terms of performance. DeepJSCC performs worse than ASCViT-JSCC at three SNRs. In real wireless channels, the state-of-the-art traditional scheme typically leads to \textit{cliff effect} when channel environments deteriorate sharply. Meanwhile, NN-based schemes not only provide graceful degradation with channel quality but also achieve results similar to or better than the state-of-the-art separation-based digital scheme.
\begin{table*}[t]
  \vspace{-3mm}
  \begin{center}
    \caption{Parameters of ICP}
    \label{tab1}
    %\resizebox{1\columnwidth}{!}{
        \begin{tabular}{|c|c|c|c|} 
          \hline
          \textbf{Carrier frequency} & 2 GHz & \textbf{System bandwidth} & 0.364 MHz \\
          \hline
          \textbf{Sampling frequency} & 1 MHz & \textbf{Subcarrier spacing} & 3.906 kHz \\
          \hline     
          \textbf{Symbols per frame} & 41 & \textbf{FFT size} & 256 \\
          \hline
          \textbf{OFDM symbol duration} & 0.32 ms & \textbf{Frame duration} & 13.12 ms \\
          \hline
        \end{tabular}
    %}
  \end{center}
  \vspace{-3mm}
\end{table*}
\begin{table*}[t]
  \begin{center}
    \caption{Performance measured in real channels}
    \label{tab2}
    %\resizebox{1\columnwidth}{!}{
        \begin{tabular}{|c|c|c|c|c|} 
          \hline
          \textbf{Distance[m]} & \textbf{\makecell[c]{ASCViT-JSCC R\\PSNR+CS/SSIM+CS}} & \textbf{\makecell[c]{ASCViT-JSCC\\PSNR+CS/SSIM+CS}} & \textbf{\makecell[c]{DeepJSCC\\PSNR+CS/SSIM+CS}} & \textbf{\makecell[c]{BPG LDPC 16QAM\\PSNR+CS/SSIM+CS}} \\
          \hline
          0.2 & 0.911/0.936 & 0.924/0.945 & 0.781/0.838 & \textbf{0.927/0.949}\\
          \hline
          1.4 & \textbf{0.837/0.845} & 0.803/0.852 & 0.639/0.764 & 0.443/0.458 \\
          \hline
          2 & \textbf{0.431/0.457} & 0.286/0.358 & 0.221/0.274 & 0/0 \\
          \hline
        \end{tabular}
    %}
  \end{center}
  \vspace{-6mm}
\end{table*}

Through the deployment and analysis of the proposed scheme and other schemes, we have verified the rationality of ICP and the advantages of ASCViT-JSCC. This establishes a solid hardware foundation for researches on intelligent communication, including semantic communication. Subsequent studies in intelligent communication can rely on ICP for practical tests to verify feasibility and advantages, and to obtain practical measurements. This will serve as a dataset source for intelligent communication and also promote the standardization of intelligent communication.

%%%%%%%%%%%%%%%%  5 CONCLUSION %%%%%%%%%%%%%%%%%%%%%%
\section{CONCLUSION}
In this paper, we have proposed a novel JSCC architecture named ASCViT-JSCC compliant with digital modulation standards for wireless image semantic transmission. We developed an adaptive preprocessing based on the object detection of YOLOv5 and feature points detection of SIFT, and designed a ViT-based JSCC network specially to preserve objects in images. Simulation results indicated our scheme can effectively preserve objects and improve the quality of reconstructed images. Additionally, we developed a SISO-OFDM testbed named ICP based on NVIDIA embedded kits and a SDR to validate intelligent communication algorithms. By deploying ASCViT-JSCC and other comparative algorithms on ICP, we verified the advantages of the proposed scheme and obtained practical measurements. Both simulations and practical measurements have verified the superiority and robustness of the proposed scheme. 

\bibliographystyle{IEEEtran} % IEEE风格
\bibliography{ref}  % BibTex文件

\end{document}